\documentclass[prd,superscriptaddress,onecolumn,showpacs,amsmath,
preprintnumbers,showkeys]{revtex4}

\usepackage{graphicx}% Include figure files
\usepackage{dcolumn}% Align table columns on decimal point
\usepackage{bm}% bold math

\usepackage{epstopdf}
\usepackage{epsfig}
\usepackage{pdfsync}
\def\beq{\begin{equation}}
\def\eeq{\end{equation}}
\def\beeq{\begin{eqnarray}}
\def\eeeq{\end{eqnarray}}

\begin{document}
\title{Hard four-jet production in pA collisions}
\author{ B.\ Blok}
\affiliation{Department of Physics, Technion---Israel Institute of
Technology, 32000 Haifa, Israel} \email{blok@physics.technion.ac.il}
\author{ M.\ Strikman}
\affiliation{Physics Department, Penn State University, University
Park, PA, USA} \email{strikman@phys.psu.edu}
\author{ U.A.\ Wiedemann}
\affiliation{Theory Division,CERN, CH-1211,Geneve, Switzerland} \email{urs.wiedemann@cern.ch}

\preprint{CERN-PH-TH/2012-253}

\begin{abstract}
In a suitably chosen back-to-back kinematics, four-jet production in hadronic collisions is known
to be dominated by contributions from two independent partonic scattering processes, thus giving
experimental access to the structure of generalized two-parton distributions ($_2$GPDs).
Here, we show that a combined measurement of the double hard four-jet cross section in proton-proton
and proton-nucleus collisions will allow one to disentangle different sources of two-parton correlations
in the proton, that cannot be disentangled with 4-jet measurements in proton-proton collisions alone.
To this end, we analyze in detail the structure of $_2$GPDs in the nucleus (A), we calculate in the independent
nucleon approximation all contributions to the double hard four-jet cross section in pA, and we determine
corrections arising from the nuclear dependence of single parton distribution functions. We then outline
an experimental strategy for determining the longitudinal two-parton correlations in the proton.
\end{abstract}

\maketitle
\setcounter{page}{1}
%\thispagestyle{empty}
%\newpage

\section{Introduction}
With increasing center of mass energy at hadron colliders, multi-particle final states at high transverse momentum receive an increasingly important contribution from multi-parton interactions. The prototype of such processes are double parton interactions in which two partons from each hadron enter into collision at two distinct hard vertexes. The calculation of such double parton interactions involves double Generalized Parton Distributions ($_2$GPDs) that - in principle - contain information about the spatial and momentum correlations of two partons in the incoming hadronic wave function. In practice, experimental constraints on these correlations are scarce. Since nuclear projectiles offer significantly different two-parton correlations,  the question arises to what extent a combined analysis program of proton-proton and proton-nucleus collisions at the LHC could help to
constrain $_2$GPDs. In this paper, we classify the different contributions to nuclear $_2$GPDs and how they contribute to the double parton scattering cross section into 4 jets that is experimentally accessible in the upcoming p-A run at the LHC.

We focus on the physics in a suitably chosen back-to-back kinematics: the dijet momentum imbalances $\delta_{13}^2,\delta_{24}^2$ are
$\delta_{13}^2=(\vec k_{1t}+\vec k_{3t})^2\ll k_{1t}^2\sim k_{3t}^2\sim Q_1^2,\delta_{24}^2=(\vec k_{2t}+\vec k_{4t})^2\ll k_{2t}^2\sim k_{4t}^2\sim Q_2^2$, where the momenta $\vec k_{1t},\vec k_{2t},\vec k_{3t},\vec k_{4t}$ are the transverse momenta of individual jets,
the scales $Q_1, Q_2$ are the resolution scales of two  partons. To justify such an ordering of momentum scales, one may want to
require $\vert k_{it}\vert$ of the order of 10 GeV or larger. 
In this kinematics the production of four jets
in the collision of two partons (so-called $2\to 4$ process) is suppressed in the leading logarithmic
approximation compared to the hard processes that involve the collision of four partons \cite{BDFS1,BDFS2,BDFS3}. For this back-to-back kinematics, it is useful to
write $_2$GPDs as the sum of two contributions,
\beq _2G_{p}
(x_1, x_2, Q^2_1,Q^2_2,\vec\Delta )
=  G_{p}^{\rm double} (x_1, x_2, Q^2_1,Q^2_2,\vec\Delta )
+ G_{p}^{\rm single}(x_1, x_2, Q^2_1,Q^2_2,\vec\Delta )\, .
\label{i1}
\eeq
Here, $_2$GPDs are written as functions of the momentum fractions $x_1, x_2$ and the resolution scales $Q_1, Q_2$ of the two partons.
In the following, we will not write explicitly the dependence of $_2$GPDs and GPDs on resolution scales.
The transverse momentum parameter $\vec \Delta$
denotes the difference between the transverse momenta exchanged by one parton in the amplitude and complex conjugate amplitude and it is conjugate to the relative spatial transverse distance
between the two partons. In general, $_2$GPDs contain a non-perturbative two-parton contribution in which QCD evolution amounts to an independent evolution of both partons
with standard one-parton evolution equations ($G_{p}^{\rm double}$). In addition, there is a contribution
in which both partons result perturbatively as the two daughters of a single parent
parton in the QCD evolution ($G_{p}^{\rm single}$). This second term couples the evolution of $_2$GPDs
to the evolution of the standard one-parton distribution functions. For the production of two
pairs of jets in independent back-to-back kinematics, both contributions are known to contribute with parametrically equal
weights~\cite{BDFS3}.

The double hard four jet cross section for the collision of hadrons $A$ and $B$ can then be written
in terms of $_2$GPDs as
\beq
\frac{d\sigma^{AB}_{4jet}}{d\hat t_1d\hat t_2}= \int\frac{ d^2\vec\Delta}{(2\pi)^2}
\frac{d\hat{\sigma}_1(x_1',x_1)}{d\hat t_1}\frac{d\hat{\sigma}_2(x_2',x_2)}{d\hat t_2}
 \, \, _2G_{A}(x_1',x_2',\vec\Delta)\, _2G_{B}(x_1,x_2,\vec\Delta)
  \, .
 \label{eq2}
 \eeq
 We denote by $\hat{\sigma}_i$ the partonic $2\to 2$ scattering
 cross sections. The measurable four-jet cross
 section in $AB$ collisions, $\sigma^{AB}_{4jet}$, is a function
of the four jets' c.m. transverse energy and rapidity, that are
connected to the variables $x_i$, $x'_i$ and the virtualities
in the standard way.
  For the
 case that $\hat{\sigma}_1$ and $\hat{\sigma}_2$ denote indistinguishable scattering processes,
 equation (\ref{eq2}) must be multiplied by a symmetry factor $1/2$ that we omit for briefness.
 In proton-proton collisions, it is customary to parametrize the cross section $\sigma^{pp}_{4jet}$
 as the product of two two-jet cross sections $\sigma^{pp}_{2jet}$,
\beq
\frac{d\sigma^{pp}_{4jet}}{d\hat t_1d\hat t_2}= \frac{1}{S} \frac{d\sigma^{pp}_{2jet}}{d\hat{t_1}}\,
 \frac{d\sigma^{pp}_{2jet}}{d\hat{t_2}}
 \, .
 \label{eq2b}
 \eeq
Here the quantity $S$ (sometimes referred to as $\sigma_{\rm eff}$) characterizes the
effective transverse area of the four parton interaction and the effect of longitudinal correlations between the partons in the colliding hadrons.
In general, $S$ can depend on the momentum fractions $x_i$ and virtualities of the incoming partons.
 Data from the Tevatron indicate that $S$ is typically of the order of 15 mb \cite{Tevatron1,Tevatron2,Tevatron3,Galina1}. This is a factor $\sim 2$ smaller than expectations
based on uncorrelated two-parton distributions, see e.g. \cite{Frankfurt}. It is a clear indication
that non-trivial parton correlations are experimentally accessible in double hard parton interactions. To date, it remains an open question
of whether these two-parton correlations in the proton are
predominantly transverse (as realized e.g. in models that picture
the proton as composed of several hot spots) or predominantly
longitudinal (as resulting e.g. from perturbative $1\to 2$ splittings).
These and other questions about the Tevatron data have
motivated much work recently~\cite{Treleani,Wiedemann,Frankfurt1,SST,Pythia,Herwig,Lund,Diehl,DiehlSchafer,Berger,Maina,Snigirev,stirling,stirling1,Ryskin,Manohar,Fano}. Older relevant work includes \cite{TreleaniPaver,mufti}.

 The case of double hard four jet production in proton-nucleus collisions was discussed first in \cite{ST}
where it was pointed out that such measurements would be sensitive to the longitudinal correlations of
the partons. More recently, multiple parton interactions were considered in the production of two leading
pions in deuteron-gold collisions at RHIC~\cite{Strikman:2010bg}, and for the case of proton-deuteron collisions~\cite{Treleani:2012zi}.
 In the present paper, we aim to extend this discussion in the light of the recent pQCD studies~\cite{BDFS1,BDFS2,BDFS3}
mentioned above.  For the case that $A$ is a nucleus, the $_2$GPD may be written as the sum of three distinct contributions, depending
 on whether both partons belong to the same nucleon (1N) or to different nucleons (2N). Distinguishing
 for the first case again single from double contributions, as in eq.~(\ref{i1}), one obtains
\beq
_2G_A(x_1,x_2,\vec \Delta)= G_A^{\rm single, 1N}(x_1,x_2,\vec \Delta)+
G_A^{\rm double, 1N}(x_1,x_2,\vec \Delta)+ G_A^{2N}(x_1, x_2,\vec \Delta
)\, .\label{sum1}
\eeq
As will become clear in the following, this classification relies on viewing the nucleus
as a superposition of nucleons to which partons can be asigned uniquely.
 Inserting the expressions for $_2$GPDs inside a nucleus (\ref{sum1}) and inside a nucleon (\ref{i1}) into
 the double hard four jet cross section (\ref{eq2}), the following different contribution emerge for the
 case of proton-nucleus scattering:
 \begin{enumerate}
  \item[I.  ]  $G_{p}^{\rm double} \otimes G_A^{\rm double, 1N}$ \\
 \item[II. ] $G_{p}^{\rm double} \otimes G_A^{\rm single, 1N}$
 \item[III.]  $G_{p}^{\rm double} \otimes G_A^{2N}\Big\vert_{\rm direct}$
 \item[IV.]  $G_{p}^{\rm single} \otimes G_A^{\rm single, 1N}$
 \item[V. ] $G_{p}^{\rm single} \otimes G_A^{\rm double, 1N}$
  \item[VI.] $G_{p}^{\rm single} \otimes G_A^{2N}\Big\vert_{\rm direct}$
 \end{enumerate}

 The terms in which two partons are taken from the same nucleon in a nucleus
 contribute equally to proton-nucleon collisions. More specifically, these are the $4 \to 4$ contribution
 (case I) and the $3\to 4$ contributions (case II and V). They will be discussed in section~\ref{sec2}.
 It is known that the single-single contribution (case IV) does not make a dominant contribution to the back-to-back kinematics; rather, it can be regarded as a one-loop correction to the $2\to 4$ cross
 section~\cite{BDFS1}.

 For a nuclear projectile, two additional contributions arise. These involve two partons from different
 nucleons in the nucleus that interact with $G_{p}^{\rm double}$ (case III) or $G_{p}^{\rm single}$ (case VI)  of the proton, respectively. These will be discussed in section~\ref{sec3}.
 We have labeled both these contributions with the subscript 'direct' to indicate that the partons with momenta $x_1$ and $x_2$ are assigned to the {\it same} nucleons in amplitude and complex conjugate amplitude.

 Viewing the nucleus as
a bound state of many nucleons  (without considering modifications to their internal structure)
 is an approximation.
 If we relax the working hypothesis that partons can be assigned uniquely to nucleons in a nucleus,
 then additional contributions are possible.   First, it is possible that two nucleons of the
 nuclear wave function are involved in both amplitude and complex conjugate amplitude, but that
 the two partons are interchanged across the cut. We label these contributions with the subscript
 'interference'. Second,  it is conceivable that the two partons taken from the nucleus belong to one
nucleon in the amplitude but to two different nucleons in the complex conjugate amplitude.
 \begin{enumerate}
   \item[VII. ]  $G_{p}^{\rm double} \otimes G_A^{2N}\Big\vert_{\rm interference}$
     \item[VIII.] $G_{p}^{\rm single} \otimes G_A^{2N}\Big\vert_{\rm interference}$
     \item[IX.  ] 1N-2N interference
 \end{enumerate}
In general, such interference terms indicate the break-down of a probabilistic
picture of double-parton interactions in pA collisions in terms of an independent superposition of
nucleons. By their very nature, they characterize partonic cross-talk between different nucleons
in a nucleus and may thus help to elucidate partonic nuclear structure.
We note that deviations from the simple picture of a nucleus as a superposition of nucleons
are known already on the level of single parton distributions, e.g. as EMC and nuclear shadowing
effects \cite{Frankfurt:2012qs}.
% Still, nuclear parton distribution functions are conveniently characterized by quantifying
% their deviations from the simple assumption of an incoherent superposition of nucleon pdfs.
% Here, we follow an analogous approach for the characterization of generalized nuclear two-parton
% distributions.
In the following sections, we discuss the different contributions to
the double-hard four jet cross section following the classification listed above.

\begin{figure}[t]  %  figure placement: here, top, bottom, or page
  % \centering
  \vspace*{-1.3cm}
   \hspace{0cm}\includegraphics[width=.70\textwidth]{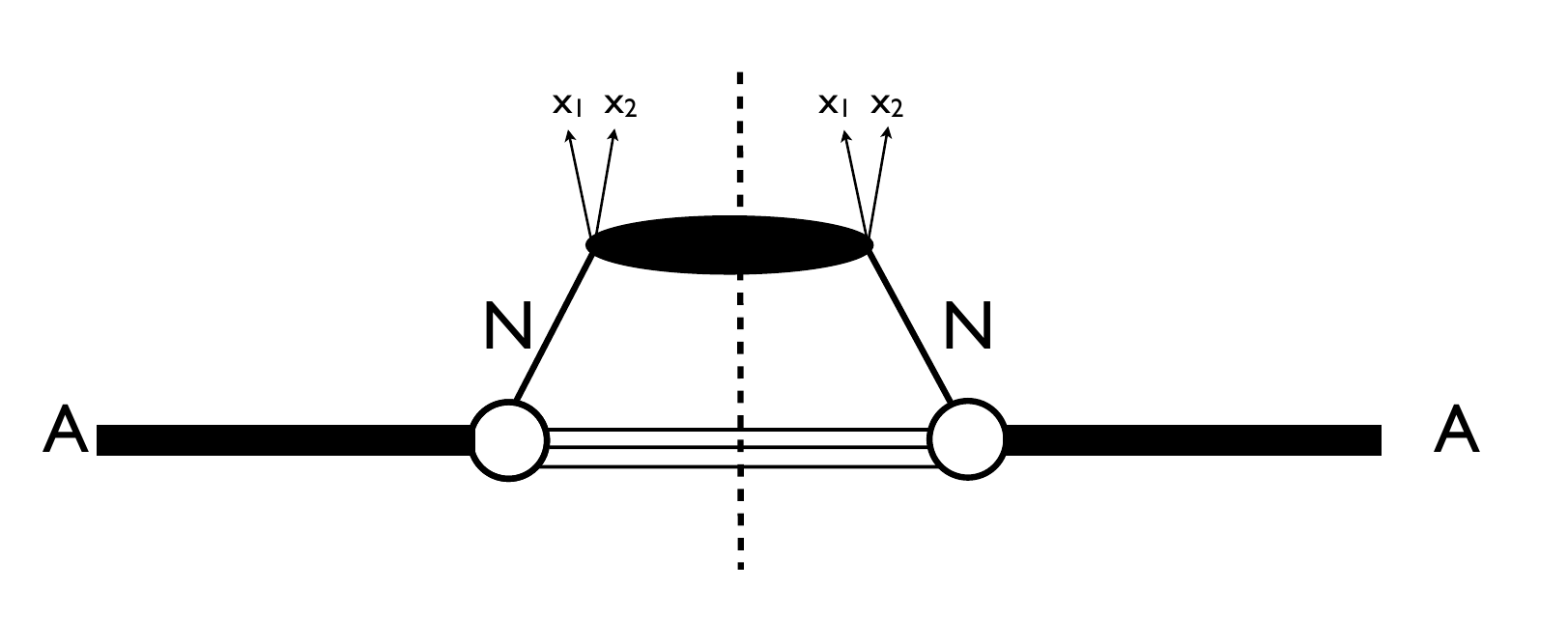}
 % \vspace*{-2.3cm}
   \caption{Schematic representation of the contributions $G_A^{\rm single,\, 1N} + G_A^{\rm double,\, 1N}$
   to the nucleus $_2$GPD that enter the cross section for double hard 4-jet production (cases I, II and IV, V
   discussed in the text). Two patrons of momentum fractions $x_1$, $x_2$ are drawn from the same single nucleon
   in the amplitude (left hand side of the diagram) and complex conjugate amplitude.
   Depending on whether the two nucleons arise perturbatively in a $1\to 2$ splitting from a single parton in
   this nucleon, or whether they are of non-perturbative origin, we shall refer to them as $G_A^{\rm single,\, 1N}$ and
   $G_A^{\rm double,\, 1N}$, respectively.}
    \label{fig1}
 \end{figure}

%%%%%%%%%%%%%%%%%%%%%%%%%%%%%%%%%%%%%%%%%%%%%%%%%%
\section{Single nucleon scattering terms (cases I, II and V)}
\label{sec2}
For the contributions I.-VI.,  the momenta of the nucleons are the same in amplitude and complex
conjugate amplitude. Therefore, if both partons belong to the same nucleon (cases I, II and V),
one can  integrate over the momenta of all other nucleons and write the corresponding part of the nuclear $_2$GPD as
\begin{equation}
 G_A^{1N}(x_1,x_2, \vec\Delta) = \int {1\over \alpha^2}   \, \,
  \left( G^{{\rm single}}_N({x_1\over \alpha},{x_2\over \alpha}, \vec\Delta)
  + G^{{\rm double}}_N({x_1\over \alpha},{x_2\over \alpha}, \vec\Delta) \right)
   \rho_A^N(\alpha, p_t) {d\alpha\over \alpha} d^2p_t \, .
\label{gimp}
\end{equation}
Here, the nuclear $_2$GPD is the sum of the terms $G_A^{single, 1N}$ and $G_A^{{\rm double}, 1N}$, introduced before, and expressed
as an integral over the corresponding contributions of individual nucleons.
The quantity $\rho_A^N(\alpha, p_t) $ denotes
the light-cone nucleon density of the nucleus normalized
as $\int \rho_A^N(\alpha, p_t) d\alpha/\alpha=A$.
The factor $1/\alpha$ for each of the partons reflects the fact that the number of partons between $x_1$ and $x_2$ should not change under Lorentz boosts.
These extra factors $1/\alpha$ are absorbed in the flux such that one recovers for the
structure function the standard expression
$ F_{2A}(x,Q^2)=  \int  \rho_A^N(\alpha, p_t) {d\alpha\over \alpha} d^2p_t F_{2N}({x\over \alpha},Q^2)$.

Equation (\ref{gimp}) is written for an ensemble of $A$ moving nucleons satisfying the
momentum sum rule
$\int \alpha \rho_A^N(\alpha, p_t) d\alpha/\alpha d^2p_t=A$ since  $\sum_i \alpha_i=A$. This raises the question of how well one can  approximate $G_A^{1N}$ in terms of the distributions
$G^{single}_N({x_1},{x_2}, \vec\Delta) + G^{{\rm double}}_N({x_1},{x_2}, \vec\Delta)$ written in the nucleus rest frame. To address this question, we replace $\alpha = 1 + (\alpha -1)$ in the arguments of
$(1/\alpha^2)\, G^{1N}({x_1\over \alpha},{x_2\over \alpha}, \vec\Delta) $, and we expand in powers of $(\alpha -1)$. Using momentum sum rule and baryon number sum rule, we find
$\int \frac{d\alpha}{\alpha} (\alpha -1) \rho_A^N(\alpha) = A - A = 0$, see Ref.~\cite{FS81,FS88}.
Therefore, corrections due to Fermi motion (i.e., corrections due to the $\alpha$-dependence of the integrand
of $ G^{1N}_A$) arise only to second order in $(\alpha -1)$. The longitudinal momentum distribution of
nucleons in a nucleus peaks at $\alpha=1$ with small dispersion, and therefore
\begin{equation}
 G_A^{1N}(x_1,x_2, \vec\Delta) =
  A\, G_N({x_1},{x_2}, \vec\Delta) \left( 1 + O\left( \int \, (\alpha-1)^2
   \rho_A^N(\alpha, p_t) {d\alpha\over \alpha} d^2p_t \right)\right) \, .
\label{gimp2}
\end{equation}
Here, the correction term involves first and second derivatives of the nucleon $_2$GPD with
respect to $x_1$ and $x_2$. A precise numerical estimate will have to constrain this term
numerically. Parametrically, the correction is small. The dominant linear dependence of
$ G_A^{1N}(x_1,x_2, \vec\Delta) $ on nucleon number $A$ translates directly into a
linear dependence of the corresponding contribution to the double hard four-jet cross section
 \begin{equation}
\frac{\sigma^{pA, 1N}_{4jet}}{d\hat{t_1}\, d\hat{t_2}} \approx
A \frac{d\sigma^{pp}_{4jet}}{d\hat{t_1}\, d\hat{t_2}} =
\frac{A}{S} \frac{d\sigma^{pp}_{2jet}}{d\hat{t_1}} \frac{d\sigma^{pp}_{2jet}}{d\hat{t_2}}
\, .
\label{singleq}
\end{equation}
We have introduced an $\approx$ sign in this relation to indicate that the identification of
$\frac{d\sigma^{pA, 1N}_{4jet}}{d\hat{t_1}\, d\hat{t_2}}$ with $A$ times the corresponding
cross section in pp relies on neglecting the nuclear modification of parton distribution functions.
Our discussion up to section~\ref{sec6} will rely on this approximation. This is justified since
we consider larger $Q^2$ processes at moderate $x$, where nuclear modifications are expected
to be small. Within this approximation, the contributions to the double hard four jet cross section in
pA discussed here are exactly the contributions that one obtains from superimposing four jet cross sections from A
independent nucleon-nucleon collisions;  the effective transverse area $S$ in (\ref{singleq})
is therefore the quantity measured in pp collisions. In section~\ref{sec7}, we go beyond this
approximation and we discuss how the nuclear dependence of parton distribution
functions can be taken into account.

 \begin{figure}[t]  %  figure placement: here, top, bottom, or page
 % \centering
  \vspace*{-1.3cm}
   \hspace{2cm}\includegraphics[width=.7\textwidth]{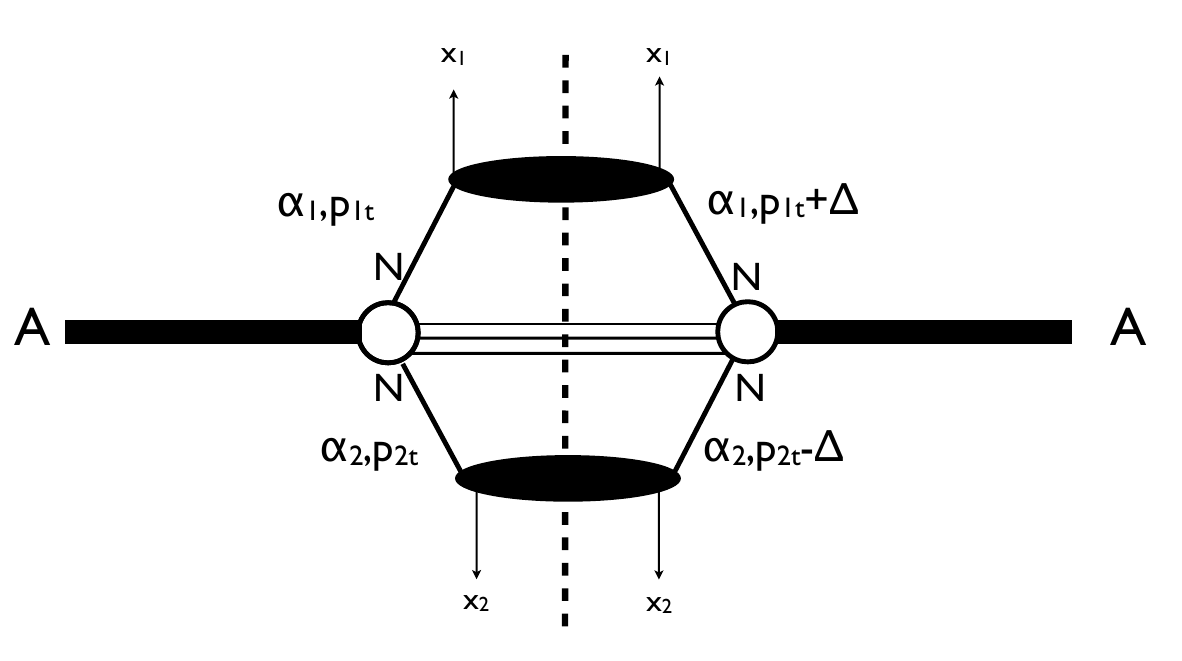}
    % \vspace*{-2.3cm}
   \caption{Schematic representation of the contribution $G_A^{\rm 2N}$ to the nucleus $_2$GPD
  that enters the cross section for the double hard 4-jet production in the terms III and VI discussed in
  the text. Here, the two patrons are drawn from two different nucleons that carry longitudinal
  nucleon momentum fractions $\alpha_1$ and $\alpha_2$. }
    \label{Fig2}
\end{figure}

%%%%%%%%%%%%%%%%%%%%%%%%%%%%%%%%%%
\section{Double nucleon scattering terms (cases III and VI) }
\label{sec3}

Figure~\ref{Fig2} shows the $_2$GPD contribution $G_A^{2N}$ in which the two partons belong
to two different nucleons in both amplitude and complex conjugate amplitude.
This term enters the contributions III and VI of the double hard four jet cross section.
In terms of the nuclear light cone wave function $\psi_A$ of the
$A$-nucleon system, it takes the form
\begin{eqnarray}
G_A^{2N}(x_1,x_2,\vec \Delta) &=& A(A-1) \int
{1\over \alpha_1 \alpha_2}
\prod_{i=1}^{i=A}\frac{d\alpha_i d^2p_{ti}}{\alpha_i}
 \, \delta\left(\sum_i \alpha_i - A\right)\,
 \delta^{(2)}\left(\sum_i {\bf p}_{ti} \right)\,
 \psi_A^*(\alpha_1, \alpha_2, p_{t1},p_{t2},....)\\ \nonumber
&& \qquad \qquad \times
\psi_A(\alpha_1, \alpha_2, p_{t1}+\vec\Delta, p_{t2}-\vec \Delta,....)
G_N(x_1/\alpha_1,\vert \vec\Delta\vert) G_N(x_2/\alpha_2,\vert \vec\Delta\vert).
\label{eq7}
\label{double}
\end{eqnarray}
Here, the transverse momentum transfer $\vec\Delta$ is exchanged between the two active nucleons, while their light cone fractions are conserved. The partonic momentum fractions $x_i$ drawn from the two active nucleons are then determined via the generalized parton distributions
$G_N(x_i/\alpha_i,\vec\Delta)$ of the nucleons. Since the wave function $\psi_A $ is normalized to unity, the prefactor $A\left(A-1 \right)$ results from combinatorics. The factor $1/\left( \alpha_1 \, \alpha_2\right)$ has the same origin as the factor $1/\alpha^2$ in eq.~(\ref{gimp}).

Expanding in equation (\ref{eq7}) the arguments of $G_N$ in powers of $(\alpha_i -1)$,
the leading term can be written in a factorized form involving the two-nucleon form factor
$F^{\rm double}_A$
\begin{eqnarray}
G_A^{2N}(x_1,x_2, \vec{\Delta}) &=& A(A-1) G_N(x_1,  {\vert\vec\Delta\vert})G_N(x_2,  \vert\vec\Delta\vert )\,\, F^{\rm double}_A( \vec{\Delta},- \vec{\Delta})\, ,
\label{eq8}\\
F_A^{\rm double}( \vec{\Delta}, -  \vec{\Delta}) &=&
\int
\prod_{i=1}^{i=A}\frac{d\alpha_i d^2p_{ti}}{\alpha_i}
 \, \delta\left(\sum_i \alpha_i - A\right)\,
 \delta^{(2)}\left(\sum_i {\bf p}_{ti} \right)
 \psi_A^*(\alpha_1, \alpha_2, p_{t1},p_{t2},....) \nonumber \\
 && \qquad \qquad \qquad \qquad \times
\psi_A(\alpha_1, \alpha_2, p_{t1}+\vec\Delta, p_{t2}-\vec \Delta,....)\, .
\label{eq9}
\end{eqnarray}
As for subleading correction, we note that one can exploit the symmetry of the
integrand of (\ref{eq7}) under $\alpha_i\to \left(2 - \alpha_i\right)$ in the nonrelativistic limit
to see that corrections to (\ref{eq8}) are of the order $(1-\alpha)^2$, similar to the case
of eq.~(\ref{gimp2}).
Since the momentum fraction $\alpha_i$ of all nucleons in the nucleus are close to unity,
we can approximate them in the nucleus rest frame in the non-relativistic limit, $\alpha_i = 1 + p_{3i}/m_N$. The two-nucleon form factor reads then
\begin{equation}
F_A^{{\it NR} \,\, {\rm double}}( \vec{\Delta}, -  \vec{\Delta})  = \int
\left( \prod_{i=1}^{i=A} d^3p_i\right)
\psi_A^*(p_1,p_2,...)\psi_A (p_1+ \vec{\Delta}, p_2 - \vec{\Delta}, p_3,...)
\delta^{(3)}\left(\sum_{i=1}^A p_i\right).
\label{eq10}
\end{equation}
Equations (\ref{eq9}) and (\ref{eq10}) express the two-nucleon form factor for an arbitrary
nucleus wave function, and therefore can account for arbitrary nucleon correlations.
Also, when combined with eq.(\ref{eq8}), these expressions can account for the finite size of nucleons as well
 (which is characterized by the single nucleon GPDs).
We now discuss how more explicit expressions, suitable for direct numerical evaluation,
can be obtained if assumptions about nucleon correlations and the finite size of nucleons
are made.

First, we turn to the independent nucleon approximation, when the nuclear wave function
is written as a product of single nucleon wave functions. This neglects all internucleon correlations, including constraints
from recoil that arise from the  kinematic $\delta$-function in (\ref{eq10}). (These latter corrections are proportional to $1/A$.
A parametrically more important source of corrections to this picture of double hard 4-jet production
arises from short-range NN interaction that are
suppressed by a factor $\propto 1/A^{1/3}$.)
One can express (\ref{eq10}) in terms
of products of Fourier transforms of single nucleon wave functions $\psi_N({r_i})$.
Using a single nucleon density $\rho_A(r) = A\,  \psi_N^*(r)\, \psi_N(r)$, that is
normalized to  $\int \rho_A(r) d^3 r= A$, the two-nucleon form factor is
the product of single nucleon form factors
\begin{equation}
F^{\rm double}_A( \vec{\Delta}, -  \vec{\Delta})
\simeq \Big\vert \int d^3r\, {1\over A} \rho_A(r)\, \exp\left[ i \vec{\Delta}\cdot \vec{r} \right] \Big\vert^2 =
F_A( \vec{\Delta})^2\, .
\label{eq11}
\end{equation}
Since $\vec{\Delta}$ is a two-dimensional vector in the transverse plane,
the single nucleon form factor can be written in terms of the nuclear thickness function
$T(b) = \int_{-\infty}^{\infty} dz \rho_A(z,\vec b)$ as
\beq
F_A(\vec \Delta )=\frac{1}{A}\int d^2b\,  T(\vec b)\exp (i \vec{\Delta}\vec b)\, .
\label{eq12}
\eeq
The well-known approximate relation between the form factor and the nucleus radius
$R_A$,
\begin{equation}
F_A^{\rm double}( \vec{\Delta}, - \vec{\Delta}) \approx \exp\left[- \frac{1}{3}\Delta^2 R_A^2\right]
\label{appr}
\end{equation}
can then be obtained by expanding (\ref{eq12}) for small $\Delta$,
$F_A(\vec \Delta ) \simeq 1 - \frac{1}{6} \Delta^2 R_A^2$ and reexponentiating this
expression. However, the Gaussian  approximation (\ref{appr}) somewhat underestimates
the drop of $F_A(\Delta^2)$  with $\Delta^2$  for $\Delta^2 R_A^2/6 \ge 1$. So, while
(\ref{appr}) is well-suited for parametric arguments, it is preferable to base numerical estimates
on evaluating (\ref{eq12}) without further approximation.

Second, we discuss now approximations that amount to neglecting the nucleon size in
comparison to the nucleus size. According to (\ref{appr}), the double nucleon scattering
contribution (\ref{eq7}) to the $_2$GPD has its main support for small values of
$\Delta^2 < O\left(3/R_A^2\right)$. Parametrically, this is a factor $A^{-2/3}$ smaller than
the range of $\Delta$-values in which a nucleon $_2$GPD has support. If one neglects
the $\Delta$-scale as being small, then the single GPDs become standard parton distributions
and the $_2$GPDs become two-parton distribution functions. In particular,
the single nucleon contribution to the $_2$GPD in (\ref{gimp2}) can be approximated as
\begin{equation}
	G_A^{1N}(x_1, x_2, \vec{\Delta}) \simeq A\, G_N(x_1, x_2, \vec{\Delta})
	  \to A\, f_N(x_1, x_2)\, ,
	  \label{eq14}
\end{equation}
where $f_N(x_1,x_2)$ is the standard two-parton distribution function. Similarly,
the double nucleon contribution to the $_2$GPD in (\ref{eq8}) can be approximated as
\begin{equation}
      G_A^{2N}(x_1,x_2, \vec{\Delta}) \to f_N(x_1)\, f_N(x_2)\,\,
      	F_A^{\rm double}( \vec{\Delta},- \vec{\Delta})\, ,
	\label{eq15}
\end{equation}
where $f_N(x)$ are standard single parton distribution functions. Simplified expressions for
the double hard four jet cross section can then be obtained by inserting equations (\ref{eq14}),
(\ref{eq15}) into (\ref{eq2}).
\begin{itemize}
\item Case III\\
We consider first the $4\to 4$ contribution  to the double hard 4-jet cross section in which the
double nucleon scattering term $G_A^{2N}$ of the nuclear $_2$GPD is paired with
$G_p^{{\rm double}}(x_1,x_2,\vec\Delta)$ in the proton. We obtain
\beq
\frac{d\sigma_4^{(III)}(x_1',x_2',x_1,x_2)}{d\hat t_1d\hat t_2} =A(A-1)
\frac{d\hat{\sigma}_1}{d\hat t_1}\frac{d\hat{\sigma}_2}{d\hat t_2}\int\frac{ d^2\vec\Delta}{(2\pi)^2}
 \, \, G_{p}^{{\rm double}}(x_1',x_2',\vec\Delta )\, f_N(x_1)\,
 f_N(x_2)\, F_A^2(\vec{\Delta})\, .
 \label{eq16a}
 \eeq
 Taking the nucleus large enough to ignore
the nucleon size, see eq.~(\ref{eq15}), one can neglect the $\vec\Delta$ dependence of $G_p(x_1,x_2,\vec\Delta )$.
Then one can write
\beq
\frac{d\sigma_4^{(III)}(x_1',x_2',x_1,x_2)}{d\hat t_1d\hat t_2} =A(A-1)
\frac{d\sigma_1}{d\hat t_1}\frac{d\sigma_2}{d\hat t_2}\int\frac{ d^2\vec\Delta}{(2\pi)^2}
 \, \, G_{p}^{\rm double}(x_1',x_2',0)\, f_N(x_1)\,
 f_N(x_2)\, F_A^2(\vec{\Delta})\, .
 \label{eq16}
 \eeq
 Here $G_{p}^{\rm double}(x_1',x_2',0) \equiv f_p(x_1',x_2')$ is the double  parton distribution function,
 and $f_N$ denotes standard nucleon pdfs. For a simple parametric estimate, the form factor
 $F_A^2(\vec{\Delta})$ can be viewed as a step function with support for $\Delta^2 < 3/R_A^2 \sim A^{-2/3}$.
 Therefore, the contribution (\ref{eq16}) is $O(A^{4/3})$ which makes it $A^{1/3}$-enhanced compared to
 all contributions discussed in section~\ref{sec2}. This can also be seen after Fourier transform to $b$-space,
 if one recalls that $T(b) \propto A^{1/3} $ for typical $b \ll R_A\sim A^{1/3}$,
\begin{equation}
\frac{\sigma_4^{(III)}(x_1',x_2',x_1,x_2)}{d\hat t_1d\hat t_2} = {f_p(x_1', x_2') \over f_p(x_1') f_p(x_2')}
\frac{d\sigma^{pp}_{2{\rm jet}}(x_1',x_1)}{d\hat t_1}\,
\frac{d\sigma^{pp}_{2{\rm jet}}(x_2',x_2)}{d\hat t_2}  {(A-1)\over A}
\underbrace{\int T^2(b)d^2b}_{\propto A^{4/3}}\, .
\label{doubleq}
\end{equation}
\par Here we expressed the four jet cross section in term of full dijet differential cross sections, defined through
hard parton cross sections as:
\beq
\frac{d\sigma^{pp}_{\rm 2jet}}{d\hat t}(x_1',x_1)=f_N(x_1)f_p(x_1')\frac{d\hat{\sigma}}{d\hat t}(x_1',x_1)\, .
\label{w1}
\eeq
Except for the $(1-1/A)$  correction term, this form of the double hard 4-jet cross section
was given first in \cite{ST}.
We note that in evaluating $\int T^2(b)d^2b$ in (\ref{doubleq}),
short-range nucleon interactions can be taken into account. For  $A\sim 200$,
the resulting corrections are on the level of a few percent~\cite{MPI08}.
  \item Case VI\\
The double nucleon scattering term $G_A^{2N}$ enters also in the $3\to 4$ contribution
to the double hard 4-jet cross section, where one parton of the proton splits into two partons
with momentum fractions $x'_1$, $x'_2$,
\begin{equation}
\frac{d\sigma_4^{(VI)}}{d\hat t_1d\hat t_2} =A(A-1)
\frac{d\sigma_1}{d\hat t_1}\frac{d\sigma_2}{d\hat t_2}\int\frac{ d^2\vec\Delta}{(2\pi)^2}
 \, \, G_{p}^{single}(x_1',x_2',0)\, f_N(x_1)\,
 f_N(x_2)\, F_A^2(\vec{\Delta}) \propto A^{4/3}\, .
\label{eq18}
\end{equation}
This term has the same parametric $A^{4/3}$-enhancement as (\ref{eq16}). However, as we
explain now, its relative weight compared to $\sigma_4^{(III)}$ is significantly smaller in $pA$ than
the corresponding relative weight in $pp$ collisions. To see this, let us recall first that in $pp$
collisions the $4\to4$ contribution involves a $\Delta$-integral over the fourth power of $F_N(\Delta)$.
This follows for instance from writing in the mean field approximation each of the two $_2$GPDs in
(\ref{eq2}) as the product of two single GPDs $_1G(x,Q^2,\vec\Delta)=f_N(x,Q^2)\, F_N(\Delta)$.
On the other hand, in proton-proton collisions the $\Delta$-integral
of the $3\to 4$ contribution involves only two powers of $F_N(\Delta)$, since
$G_p^{single}(x_1,x_2,\vec\Delta)$ corresponds to a point-like perturbative splitting and
its $\Delta$-dependence is  thus negligible~\cite{BDFS1}. In general, since $F_N(\Delta)$ peaks
at $\Delta = 0$ and falls off steeply with increasing $\Delta$, the $\Delta$-integral over $F_N^2(\Delta)$
is larger than that over $F_N^4(\Delta)$. This results in a geometrical enhancement of the
$3\to 4$ contribution in $pp$ relative to the $4\to 4$ contribution. For a numerical estimate of this
enhancement in $pp$, one may take recourse e.g. to the
$F(\Delta)=\frac{1}{(1+\Delta^2/m^2_g)^2}$\cite{Frankfurt:2002ka}, for which
\beq
\int \frac{d^2\Delta}{(2\pi)^2}F_N^2(\Delta)
\Big/  \int \frac{d^2\Delta}{(2\pi)^2}F_N^4(\Delta)
=  \frac{m^2_g}{12\pi^2}  \Big/     \frac{m^2_g}{28\pi^2} = 7/3\, .
\label{w4}
\eeq
We note that this ratio is rather robust against changes of the functional shape of $F(\Delta)$; for instance,
an exponential form of $F(\Delta)$ would yield a ratio $2$ rather than $7/3$.
%
% . For the later case one would get an enhancement factor of 2 which differs  from eq.(\ref{w4}) only by 15\%. So the magnitude of this enhancement is not sensitive to the current uncertainties of the data.
%While the precise numerical value of this ratio clearly depends on model-dependent assumptions about
%the shape of $F(\Delta)$, a smaller enhancement factor would imply a weaker $\Delta$-dependence of
%$F(\Delta)$ which seems difficult to justify. We therefore regard (\ref{w4}) as indicative of a lower bound onthis ratio.
%%ms exp give smaller value so this is not a lower bound
 In contrast to this geometrical enhancement in $pp$, we have found here that in $pA$ collisions
in the limit of very large A, the $\Delta$-integrals in (\ref{eq16}) and (\ref{eq18}) are the same and a
geometrical enhancement as in (\ref{w4}) is missing. In addition, the single parton in a $3\to 4$ process
could belong to each of the colliding protons in $pp$, whereas this combinatorial factor $2$ is obviously
absent in $pA$. In summary, there is a geometrical enhancement factor of $3\to 4$
relative to $4 \to 4$ processes in $pp$ that equals $7/3\times 2=14/3\sim 5$ and that is clearly
 absent in pA collisions for sufficiently large $A$,
\beq
{\frac{\sigma_4^{(VI)}}{ \sigma_4^{(III)}} \Bigg\vert_{pA} \over
 \frac{\sigma_4^{(VI)}}{ \sigma_4^{(III)}} \Bigg\vert_{pp} }= {\rm const} (A)_{\left| A\gg 1\right.} \sim \frac{1}{5}\, .
\eeq
\end{itemize}
%
%%%%%%%%%%%%%%%%%%%%%%%%%%%%%%%%%%%%%%%%%%%%%%%%%%%%%
\section{Total cross section in independent nucleon approximation.}
\label{sec4}
\par In summary, working in the independent nucleon approximation and neglecting the nucleon size compared
to the nuclear radius $R_A$, one can write the double hard four-jet cross section in $pA$ as the sum of terms
that are linear in $A$ (see discussion of $\sigma_4^{1N}$ in section~\ref{sec2} and eq. (\ref{singleq}) ) and of
double scattering terms that were considered in section~\ref{sec3} (see eqs.(\ref{doubleq}), (\ref{eq18}) ).
We define $\sigma_{4\, jet}^{(pA)}$ as the sum of all these contribution.
Under mild assumptions, the relative weight of the $(3\to 4)$ and
$(4 \to 4)$ contributions is expected to change between $\sigma_{4\, jet}^{(pp)}$ and $\sigma_{4\, jet}^{(pA)}$, see section~\ref{sec3}. However, the analysis proposed in the present section will not depend on the numerical estimates given in section~\ref{sec3}.  We start from
the 4-jet nuclear modification factor that is constructed by
normalizing $\sigma_{4\, jet}^{(pA)}$ by $A$ times
the corresponding cross section in a nucleon-nucleon collision,
\begin{eqnarray}
R_{pA}^{4jet}(x_1,x_2,x'_1,x'_2) &\equiv& \frac{d\sigma_{4jet}^{pA}(x_1,x_2,x'_1,x'_2) }{d\hat{t}_1\, d\hat{t}_2}\Bigg/
\frac{A}{S} \frac{d\sigma_{2\, jet}(x'_1,x_1)}{d\hat{t}_1} \frac{d\sigma_{2\, jet}(x'_2,x_2)}{d\hat{t}_2}
\nonumber \\
&=& 1 + \frac{S}{A} \frac{A-1}{A}\int T^2(b) d^2b {G_p(x_1',x_2')\over  f_p(x_1')  f_p(x_2')},
\label{eq23}
\end{eqnarray}
where
\beq
G_p(x_1',x_2')=f_p(x_1',x_2')+G^{\rm single}_p(x_1',x_2',0)\, .
\label{eq24}
\eeq
We note that a closely related
expression was obtained already in the analysis of Ref.~\cite{ST}, where the $3\to 4$ contribution was
not included and $1/A$-corrections were neglected.
It is useful to write (\ref{eq23}) in the compact form

\begin{eqnarray}
R_{pA}^{4jet}(x_1,x_2,x'_1,x'_2)
= 1 + S\, W(A)\, K(x_1',x_2')\, ,
\label{eq25}
\end{eqnarray}
where the second term on the right hand side of (\ref{eq23}) factorizes into a product of
the effective transverse area $S$, a purely geometrical overlap factor
 \beq
 W(A)=\frac{A-1}{A^2}\int d^2bT^2(b)\, ,
 \label{eq26}
 \eeq
and the normalized longitudinal parton correlation function
 \begin{equation}
K(x_1',x_2')=
{G_{p}(x_1',x_2',0)\over  f_p(x_1')  f_p(x_2')}\, .
\label{eq27}
\end{equation}
The second term $S\, W(A)\, K(x_1',x_2')$ on the right hand side of (\ref{eq25}) corresponds to
the parametrically $A^{4/3}$-enhanced contributions to the 4-jet double scattering cross section that we have
discussed in section~\ref{sec3}.  For a lead nucleus with Wood-Saxon density profile and $S = 15$ mb,
one finds  $S\, W(A) \sim 2.1$~\cite{Frankfurt:2002jr}.
  This shows that these parametrically enhanced
terms give indeed the largest contribution for sufficiently heavy nuclei. However, terms with linear
$A$-dependence must not be neglected for numerical estimates, since they constitute about $1/3$
of the four-jet cross section.

Equation (\ref{eq23}) or (\ref{eq25}) summarizes one of the main results of this paper. It demonstrates
that any $x'_1$- and $x'_2$-dependence of the nuclear modification factor provides information about
the longitudinal correlation (\ref{eq24}) of two partons in the nucleon, since it constrains directly the ratio $K$.
It is in this sense that the nucleus provides a non-trivial filter for analyzing the multi-parton structure of the proton.
We further note that the effective area $S$ is operationally defined as the ratio of the double hard 4-jet
cross section and the product of two dijet cross sections in $pp$, and thus it can be a function of
$S = S(x_1,x_2,x'_1,x'_2)$. However,
 \begin{equation}
K(x_1',x_2')= \frac{R_{pA}^{4jet}(x_1,x_2,x'_1,x'_2) - 1}{S\, W(A)}
\label{eq28}
\end{equation}
can depend only on the momentum fractions $x'_1$, $x'_2$ in the proton. Any deviation of the ratio (\ref{eq28})
from unity would be an unambiguous signal of longitudinal momentum correlations in the proton.
We note that the numerical analysis of \cite{BDFS3}, based on dominance of $3 \to 4$ processes,
suggests $K\sim 1.2$ in the kinematic region under consideration.
Vice versa, since the $3\to 4$ contributions involve partonic $1\to 2$ splittings in the nucleon and thus
lead dynamically to longitudinal momentum correlations, any tight bound on
$\vert K(x_1',x_2') - 1\vert$ will put significant constraints on the role of $3\to 4$ processes.
In the extreme case when $K(x_1',x_2') = 1$, one would have to conclude that the dynamical origin of
the anomalously small effective area $S$ is purely transverse. We argue that these considerations
motivate an experimental study of the nuclear modification factor (\ref{eq23}) and the corresponding
longitudinal correlation function (\ref{eq28}) in the upcoming pA run at the LHC.
%

 %%%%%%%%%%%%%%%%%%%%%%%%%%%%%%%%%%
\begin{figure}[t]  %  figure placement: here, top, bottom, or page
  % \centering
  \vspace*{-1.3cm}
   \hspace{2cm}\includegraphics[width=.70\textwidth]{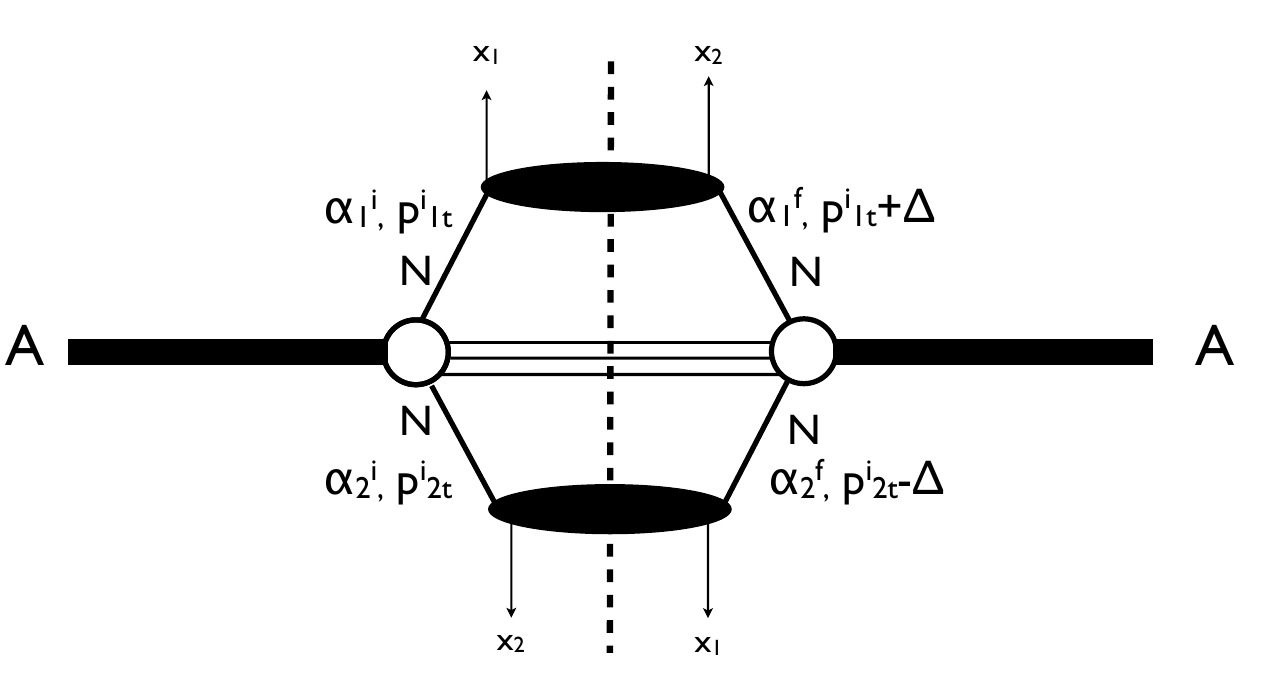}
     %\vspace*{-2.3cm}
   \caption{Schematic representation of an interference contribution to the nucleus $_2$GPD
   in which the two patrons of momentum fractions $x_1$, $x_2$ are drawn from two different nucleons
   in the amplitude, but in which the assignment between partons and active nucleons is swapped in the
   complex conjugate amplitude. As discussed in section ~\ref{sec5}, these contributions (cases VII and VIII)
   are expected to suppressed due to constraints on the longitudinal light-cone fractions of the two active nucleons.}
    \label{Fig3}
 \end{figure}
 %%%%%%%%%%%%%%%%%%%%%%%%%%%
%%

We note as an aside that the corresponding result for proton-deuteron scattering can be calculated, based on the explicit
form of the two-body deuteron form factor $F_D^{\rm double}(\vec\Delta,-\vec\Delta)$. Since the nucleons in
the deuteron are strongly correlated in the center of mass frame, $F_D^{\rm double}$ is simply expressed
through the one body deuteron form factor $F_{D}$ as
$F_D^{\rm double}(\vec{\Delta}, -\vec{\Delta}) = F_D(4\Delta^2)$,
 similar to the case of the Glauber scattering, cf.  \cite{Gribov70}.
 Inserting this expression into (\ref{eq18}) and performing the $\Delta$-integral, one obtains
 $R_{pD}^{4jet} \simeq 1 + 1.07 K$ for $x_i = O(0.01)$. We thank V. Guzey for performing this
 numerical integration and communicating the result. The case of pD scattering was considered
 recently in \cite{Treleani:2012zi} in coordinate space formalism, though no numerical results were reported.
 %%

%%%%%%%%%%%%%%%%%%%%%%%%%%%%%%%%%%%%%%%%
\section{Double scattering -- Interference term (Cases VII and VIII)}
\label{sec5}

Our discussion so far was based on a picture in which the nucleus is viewed as an independent superposition
of nucleons, even if the parton distribution functions inside each nucleon may differ from those in a free nucleon.
If one does not assume this picture, further interference contributions to the double hard four-jet cross section
are conceivable. Here, we discuss the form of these contributions for completeness. We note at the beginning
that we do not have a complete framework for calculating them, but we shall give some arguments of why we
expect them to provide at best very small corrections to the expression (\ref{eq23}).

As depicted in Fig. \ref{Fig3}, it is possible to write down a diagram where in the initial state a parton with $x_1$ 
from the nucleon "1" is involved in the two-to-two process while this nucleon absorbs the parton with $x_2$ in the final
state. This process changes the longitudinal light cone fractions $\alpha_{1/2}^{i/f}$ of the two active nucleons
between the initial and final state. In the parton model, we have
\begin{equation}
\alpha^{i}_1 - x_1+x_2=\alpha^{f}_1; \,  \alpha^{i}_2 + x_1-x_2=\alpha^{f}_2.
\end{equation}
This implies that the momentum transfer $\vec{\Delta}$ now has also a nonzero longitudinal component,
which in the nonrelativistic approximation can be written as
 \begin{equation}
\vec{\Delta} = (\Delta_3 = (x_1-x_2)m_N, \vec{\Delta}_t).
  \end{equation}
Consequently, in close analogy to the discussion of (\ref{appr}), the two-nucleon form factor takes
now the form
 \begin{equation}
 F_A( \vec\Delta, -\vec\Delta) \approx
  F_A^2((x_1-x_2)^2m_N^2 + \vec{\Delta}_t^2) \approx \exp(-((x_1-x_2)^2m_N^2 + \vec{\Delta}_t^2)R_A^2/3)
% F_A^2(\vec \Delta_t^2).
\label{eq31}
 \end{equation}
in the mean field approximation. Here, the transverse factor is of the form of (\ref{appr}). The additional
suppression factor  $\exp(-  (x_1-x_2)^2m_N^2 \cdot R_A^2/3)$ arises for significant differences in the
longitudinal momentum fractions. For a typical value of the nuclear radius at large $A$, $R_A \sim 6$ fm,
one finds a strong suppression in the range $|x_1-x_2| \ge 0.03$.
For $|x_1-x_2| \le 0.03$, this suppression factor in (\ref{eq31})  is less important and it vanishes for $x_1=x_2$. 
We emphasize, however, that the estimate (\ref{eq31}) does not account for all physics effects. In particular, it
neglects effects from QCD evolution. In the remainder of this section, we present arguments for why the contribution 
of Fig.\ref{Fig3} can be expected to be suppressed in the range $|x_1-x_2| \le 0.03$, too.

As emphasized already in the introduction, the discussion in the present paper focusses on sufficiently hard
processes for which a perturbative hierarchy of scales $\delta_{13}^2 = (\vec k_{1t}+\vec k_{3t})^2\ll k_{1t}^2\sim k_{3t}^2\sim Q_1^2$
ensures that one can select experimentally the relevant back-to-back kinematics in which double-hard four jet production
dominates. For the case of LHC, this implies that one realistically should consider jets with $k_{t} >  O(10)$  GeV/c 
and hence typically $x_i\ge 0.005$ for production at central rapidities. In general, the initial state QCD parton showers associated
to such processes lead to radiation above some non-perturbative starting scale $Q_0$ and up to the transverse momentum $k_t$.
(For the processes under consideration at the LHC, transverse momenta arising from this QCD evolution may be estimated to be 
of the order of $2$ GeV or larger for each of the two hard processes.) As a consequence of this QCD evolution, the momentum transferred 
to the two nucleons involved in Fig.\ref{Fig3} are $ \sim \pm  p^i_{1t} - p^f_{2t}$, and  $ \left|p^i_{1t} - p^f_{2t}\right|\gg Q_0$  leading to a configuration with two nucleons with back to back momenta $ p^i_{1t} - p^f_{2t} $. These momentum differences are much larger than 
the typical momenta in the nucleon wave function: $\le p_F\sim \mbox{250 MeV/c}$, but for a physical contribution, the transverse
momenta of the nucleons must match between amplitude and complex conjugate amplitude in Fig.\ref{Fig3}. This is
only possible if the initial state radiations of partons in the two active nucleons are matched to an extent that $ p^i_{1t} - p^f_{2t} $
is much smaller than what one expects from two independent QCD evolutions. This is a phase space constraint on the QCD evolution 
that is not included in the parton model estimate (\ref{eq31}). Since one requires $\vert p^i_{1t} - p^f_{2t} \vert \simeq p_F \ll \vert p^i_{1t} \vert,
\quad \vert p^f_{2t} \vert$, we expect this phase space constraint to provide a very strong suppression factor for the contribution Fig.~\ref{Fig3}. 

% In this case  one needs to take into account effects of the QCD evolution. One effect is that nucleon pdfs at virtuality $p_t^2$ and say $x\sim 0.01$ are determined by nucleon pdfs at the initial scale at much larger $x\sim 0.1$, see for example curves for the gluon evolution in Fig. 10  \cite{FGS1}  and  Fig.61 in \cite{FGS2} for gluons (for the sea (anti)quarks evolution is similar).

In the previous paragraph, we have argued that effects of QCD evolution that are not taken into account in (\ref{eq31}), lead to a strong
suppression of Fig.~\ref{Fig3}. This suppression is expected to increase with increasing jet energy when effects of QCD evolution 
become more important, i.e., this suppression is particularly relevant in the region of high transverse momentum on which our 
discussion focusses in this paper. We note that at sufficiently high jet energy, further suppression effects may arise. In particular, 
we observe that for $x_i\ge 0.05$, the coherence lengths  $\sim 1/2m_Nx_i$ become smaller than the average internucleon distance 
$r_{NN}\sim $ 2 fm. If one expects that the exchange of partons between two nucleons is similar to the diagram of NN interaction which in t-channel has the closest singularity at $m_{\pi}^2$ reflecting the finite range of NN interactions, one must require that the active nucleons 
in the nucleus are close in configurational space: $r_{NN}\le m_{\pi}^{-1}$. One arrives at the same argument by noticing 
that since in this case the coherence lengths (Ioffe times or equivalently current correlators) are small,  interference  is possible only if the longitudinal distance between nucleons is smaller than coherence length. Hence the interference term for $x_i\ge 0.05$ is not enhanced by a factor $A^{1/3}$ as the double nucleon scattering term. 
% Since the exchange diagrams can contribute with comparable strength to the nucleus pdfs  for $x\sim 0.1$ where nuclear pdfs
% are close to nucleon pdfs with accuracy of few percent, correction to the impulse approximation in the parton model should not exceed 5\%.

In summary, very little is known so far about interference contributions of the form Fig.~\ref{Fig3}.~\footnote{Interference effects were 
also considered recently in \cite{Treleani:2012zi}. In contrast to our work, however, the main focus in~\cite{Treleani:2012zi} was on the 
case of scattering of protons off the lightest nuclei ($A=2, 3$). The  A-dependence for large $A$, and the effects of QCD evolution 
suppressing interference discussed here were not addressed in  \cite{Treleani:2012zi}. }
 In the present section, we have considered this contribution first on the level of the parton model, see discussion of eq.~\ref{eq31}. 
 We have then given a qualitative argument for why a very strong additional suppression of  Fig.~\ref{Fig3}, not seen in the estimate 
 (\ref{eq31}), should arise if effects of QCD evolution are taken into account for sufficiently hard processes. And we have given a second, independent formation time argument for why the contribution Fig.~\ref{Fig3} is not enhanced by $A^{1/3}$. Both qualitative arguments 
 indicate that Fig.~\ref{Fig3} is strongly suppressed compared to the other contributions to the double hard 4-jet cross section in 
 proton-nucleus collisions. 

% {\bf We would prefer to remove the text below MS, BB}
%  {\it  Here, we do not attempt to estimate such potential further suppression effects further. We merely observe from (\ref{eq31}) that such an additional contribution, if not negligible, will correct our main result (\ref{eq25}),
%
%   \begin{equation}
%   R_{pA}^{4jet}(x_1,x_2,x'_1,x'_2) = 1 + S\, W(A)\, K(x'_1,x'_2)\, \left(1 + F_{\rm corr} \right)\, ,
%  \label{main}
% \end{equation}
%
% where the correction factor $F_{\rm corr}$ is expected to be small. In the case considered in (\ref{eq31}), this correction factor will take the explicit form $F_{\rm corr} = {\rm const}\cdot \exp(-(x_1-x_2)^2m_N^2 R_A^2/3)$, where the proportionality factor is much smaller than unity. }

\begin{figure}[t]  %  figure placement: here, top, bottom, or page
  % \centering
\vspace*{-.3cm}
 \hspace{0.1cm}\includegraphics[width=.65\textwidth]{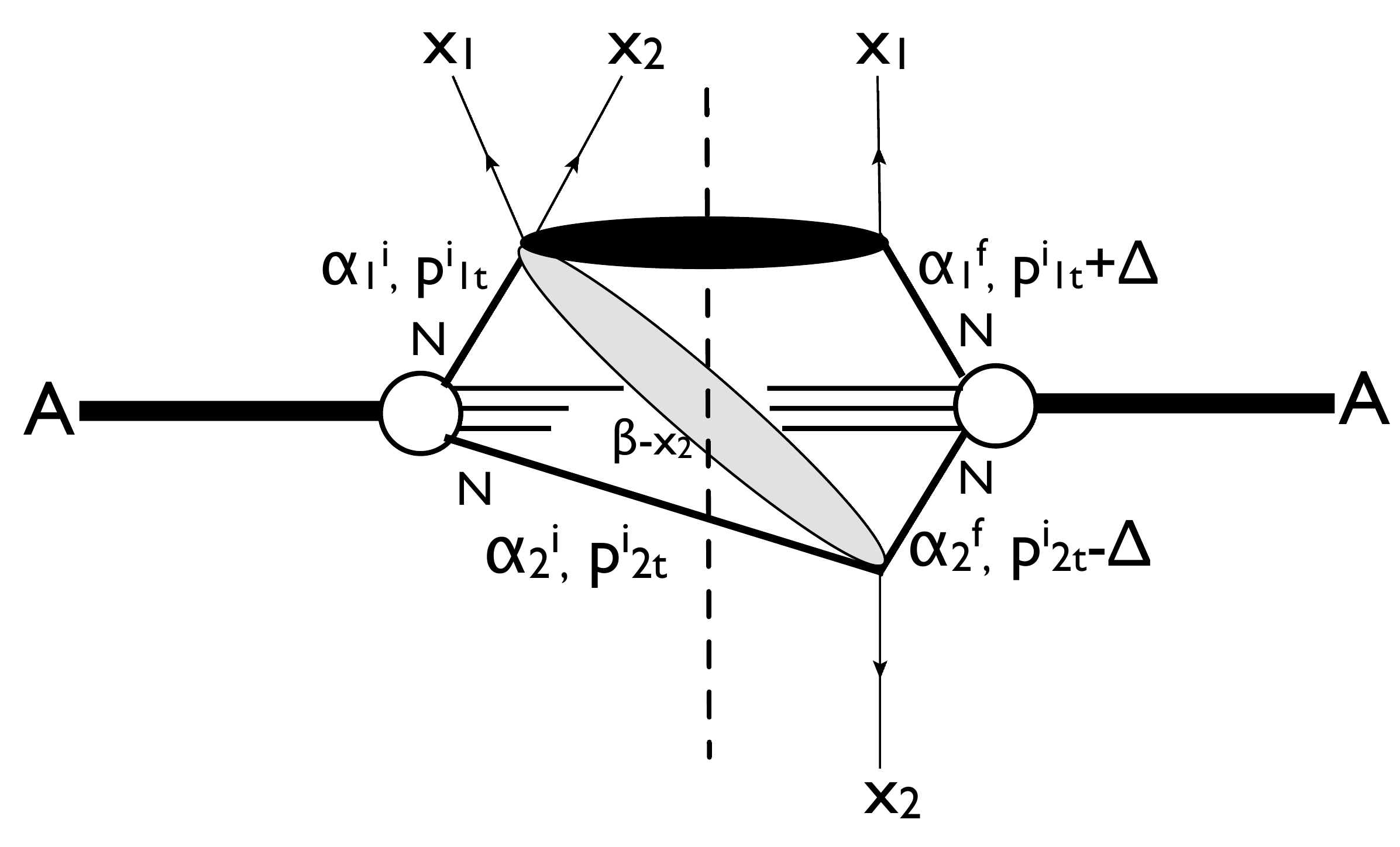}%{inter.pdf}
 % \vspace*{-1.3cm}
   \caption{Schematic representation of an interference contribution
   to the nucleus $_2$GPD in which the two partons are drawn from
   a single nucleon in the amplitude while the parton '2' is associated to
   a different nucleon in the complex conjugate amplitude. This process requires the exchange of additional constituents between both active nucleons, as indicated
   by the grey blob.}
    \label{inter}
 \end{figure}
%%%%%%%%%%%%%%%%%%%%%%%%%%%%%%%%%
\section{Two nucleon -- to -- one nucleon interference (Case IX)}
\label{sec6}

We finally consider the process where the two partons active in the double hard scattering
are drawn from the same nucleon (with light-cone momentum fraction $\alpha_1$) in the amplitude,
but from two different nucleons (with light cone momentum fractions $\alpha_3,\alpha_4$) in the
complex conjugate amplitude, see Fig.\ref{inter}. The momentum fraction of the spectator nucleon
in the amplitude is denoted by $\alpha_2$. Longitudinal momentum conservation implies constraints
on such processes. In particular, $\alpha_2 + \beta = \alpha_4$ and $x_2 \leq \beta$. Since the
light cone momentum fractions $\alpha_i$ deviate from unity only by Fermi motion, this will constrain
contributions of the diagram Fig.\ref{inter} to relatively small momentum fractions $x_2$. In addition,
we expect that exchanges as depicted in Fig.\ref{inter} can arise only between nucleons neighboring in impact parameter.
We note as an aside that the diagram Fig.\ref{inter} is related to the
diagrams for the leading twist nuclear shadowing discussed in Ref.~\cite{Frankfurt:2011cs}.
% Qualitatively, the limitations
% to sufficiently small $x$ and a further suppression of geometrical origin are the same two suppression
% factors that we identified for the cases VII and VIII discussed above.
 But to become quantitative is difficult and lies outside the scope of the present work. We now
 turn to a more general discussion of the corrections to (\ref{eq23}) that may arise from the nuclear
 dependence of parton distribution functions.

%%%%%%%%%%%%%%%%%%%%%%%%%%%%%%%%%
\section{Nuclear dependence of single parton distribution functions}
\label{sec7}

In our discussion so far,  we have neglected nuclear modifications of the single
parton distribution functions. Standard parametrizations of these modifications are based
on a linear relation between the parton distributions $f_{i/p}(x,Q^2)$ in a proton and the
parton distributions $f_{i/A}(x,Q^2)$ per nucleon in a
nucleus~\cite{Eskola:1998df,Eskola:2009uj,Hirai:2007sx,deFlorian:2003qf,Frankfurt:2011cs}
\begin{equation}
    f_{i/A}(x,Q^2) = R_i^A(x,Q^2)\, f_{i/p}(x,Q^2)\, .
    \label{eq6b}
\end{equation}
 In all parametrizations that are consistent with linear
$Q^2$-evolution~\cite{Eskola:1998df,Eskola:2009uj,Hirai:2007sx,deFlorian:2003qf,Frankfurt:2011cs}, the npdf
factors $R_i^A(x,Q^2)$ at fixed $x$ approach unity with increasing $Q^2$. Since the jet production considered
here is a hard process with typically $Q^2 \gg 100\, {\rm GeV}^2$, we expect $\vert R_i^A(x,Q^2) - 1\vert < 3\%$
over the entire $x$, and $Q^2$-range relevant for four-jet production at the LHC. This assumption can be
verified experimentally by checking that the nuclear modification factor for dijet production,
\begin{equation}
	R_{pA}^{\rm 2jet} \equiv
	  \frac{d\sigma^{pA}_{2jet}}{d\hat{t_1}} \Big /
		 A\, \frac{d\sigma^{pp}_{2jet}}{d\hat{t_2}} \, ,
		 \label{eq35}
\end{equation}
deviates from unity by less than 3\% in the kinematical range used to measure the four-jet cross sections.
An O(3\%) uncertainty from ``npdf corrections'' to the dijet cross section entering the norm in (\ref{eq23})
is expected to result in an O(6\%) uncertainty on the nuclear modification factor for double hard four-jet production.
The central question is whether a correction to (\ref{eq23}) of this order is small enough to be neglected
in the analysis of the longitudinal two-parton correlation function  $K(x_1,x_2)$. This
depends on how the size of the deviation of $K(x_1,x_2)$ from unity compares to the size of the
npdf correction of (\ref{eq23}), and we shall distinguish below two different cases.

Let us discuss, however, first how the nuclear modification of longitudinal parton momenta in the nucleus
can be taken into account in the formulation of $_2$GPDs. To this end, one needs to specify correction factors
for the three contributions to the nuclear $_2$GPD listed on the right hand side of equation (\ref{sum1}).
In principle, this requires more information than what is contained in the npdf-fits based on (\ref{eq6b}).
Adopting the picture of the nucleus as an incoherent superposition of nucleons that have parton distributions
shifted according to (\ref{eq6b}), one arrives at corrections of the simple form
\begin{eqnarray}
 	G_A^{2N}(x_1,x_2,0) &\longrightarrow& R_i^A(x_1,Q^2)\, R_j^A(x_2,Q^2)\, G_A^{2N}(x_1,x_2,0) \, ,
%	f_{i/p}(x_1,Q^2)\, f_{j/p}(x_2,Q^2)\, ,
 	\label{eq36}\\
 	G_A^{{\rm double}, 1N}(x_1,x_2,0) &\longrightarrow& R_i^A(x_1,Q^2)\, R_j^A(x_2,Q^2)\,
	G_A^{{\rm double}, 1N}(x_1,x_2,0) \, .
%	G_p^{\rm double}(x_1,x_2,0)\, .
 	\label{eq37}
\end{eqnarray}
In principle, nuclear pdfs depend also on the impact parameter $b$ (see Ref.~\cite{Frankfurt:2011cs,Helenius:2012wd} for first parametrizations), and this $b$-dependence could be accommodated in (\ref{eq36})
and (\ref{eq37}). This, however, will be a small correction on top of the correction discussed here, and
we neglect it in the following.

The third contribution $G_A^{single, 1N}$ to the nuclear $_2$GPD does not have an npdf correction factor
of this form. Rather, since the arguments $x_1$, $x_2$ in $G_A^{single, 1N}$ result dynamically from the
splitting of a single parton, the npdf corrections will be determined by an integration over the available phase
space that may be written formally as
\begin{equation}
	G_A^{single, 1N}(x_1,x_2,0) \longrightarrow O\left(R_i^A(x_1+x_2,Q^2)\right)\, G_A^{single, 1N}(x_1,x_2,0)\, .
%	G_p^{single}(x_1,x_2,0)\, .
	\label{eq38}
\end{equation}
To determine npdf corrections to (\ref{eq23}) in the most general case, one would have to specify (\ref{eq38})
fully and then repeat the calculations in sections~\ref{sec2}-\ref{sec4} based on equations (\ref{eq36})-(\ref{eq38}).
Because of the more complicated form of (\ref{eq38}), the explicit results for the npdf corrections to (\ref{eq23})
would be relatively involved. Here, we restrict our discussion by considering two limiting cases:

First, we observe that if (\ref{eq38}) makes a numerically important contribution to the total nuclear $_2$GPD
(\ref{sum1}), then longitudinal two-parton correlations in the nucleus are expected to be large. This is so,
since (\ref{eq38}) is dynamically generated by a perturbative parton branching that inevitably results in
significant longitudinal correlations. Indeed, according to the model of \cite{BDFS3} one can expect a
$\sim 20\%$ deviation of $K(x_1,x_2)$ from unity if the double hard four jet cross section is dominated
by $3\to 4$ processes. Therefore, if the term (\ref{eq38}) is relevant, we expect that $K(x_1,x_2)-1 > 0.1$
and that the npdf-corrections to (\ref{eq23}) are small. In this case, $K(x_1,x_2)-1$ is much larger than the \
expected npdf corrections and it can thus be extracted safely from (\ref{eq28}) without taking npdf corrections into
account. We emphasize that the validity of this procedure can be checked experimentally by measuring the
nuclear modification factor for dijet production in pA.

Alternatively, if an experimental determination of $K(x_1,x_2)-1$ from (\ref{eq28}) yields values
$K(x_1,x_2)-1 \ll 0.1$, the contribution of the $1\to 2$ splitting term (\ref{eq38}) to the nuclear $_2$GPD
can be expected to be small, and one can justify hence the use of a simplified npdf correction of the form
$_2G_A(x_1,x_2) \to  R_i^A(x_1,Q^2)\, R_j^A(x_2,Q^2) _2G_A(x_1,x_2) $. In this case, the npdf corrections to
the contributions (\ref{singleq}), (\ref{doubleq}) and (\ref{eq18}) amount to multiplying all three expressions
with two powers of the nuclear modification factor for dijets, $R_{pA}^{2jet}$. As a consequence, the right
hand side of (\ref{eq23}) gets multiplied by two powers of $R_{pA}^{2jet}$, and the npdf correction to
(\ref{eq28}) takes the form
 \begin{equation}
K(x_1',x_2')\vert_{\rm npdf\, corrected}= \frac{R_{pA}^{4jet}(x_1,x_2,x'_1,x'_2)/ (R_{pA}^{2jet})^2 - 1}{S\, W(A)}\, .
\label{eq39}
\end{equation}
Experimentally, comparing the values obtained from (\ref{eq28}) and (\ref{eq39})  provides a direct way
of estimating the importance of npdf corrections on the interpretation of $K$ as a normalized longitudinal
two-parton correlation function.

%%%%%%%%%%%%%%%%%%%%%%%%%%%%%%%%%%%%%%%%%%%%%%%
\section{Conclusions.}
\label{sec8}

\par In summary, by analyzing $_2$GPDS for the nucleus in the many nucleon approximation, we have
derived a compact expression (\ref{eq23}) for the nuclear modification factor of the double-hard four
jet cross section in pA collisions. Based on this main result, we have outlined an experimental strategy
for determining the normalized longitudinal two-parton correlation function $K(x_1,x_2)$ in the proton
by combining data from pp and pA collisions.
We also argued that interference contribution due to exchange of two partons between nucleons are strongly suppressed in the LHC kinematics.
 Finally, we have discussed how nuclear modifications
of parton distribution functions can be taken into account in this analysis. 
Overall our treatment allows to consider the bulk of the LHC kinematics when at least one of the nuclear partons has $x\ge 0.005$. The kinematics where both nuclear partons
are in the shadowing region will be considered elsewhere.

As emphasized in Ref.~\cite{BDFS1} and as recalled here, one possible dynamical source of longitudinal 2-parton correlations $K(x_1',x_2')$ in
the proton is collinear parton splitting that leads to a sizable $3\to 4$ contribution in 4-jet events. The comparison of data from pp and pA
collisions, advocated here, should be regarded as one of several experimentally feasible avenues to test for such a contribution. Another possibility to discriminate between $3\to 4$ and $4\to 4$ contributions
may be given by exploiting their different dependence on $\sqrt{s}$.
Given the complexity of the problem of characterizing 2GPDs, we believe
that all possible approaches should be explored. The main purpose
of this paper is to discuss how data from pA collisions can contribute to such a program.

In early 2013, the LHC is  scheduled for a 4-week-long proton-nucleus run. The main
motivation for this pA program at the LHC is to constrain the parton distributions in the nucleus and
to provide important benchmark measurements for the LHC heavy ion programme. As illustrated by the
calculation of $K(x_1',x_2')$ in this paper, however, proton-nucleus collisions may also contribute to
further constrain the multi-parton structure of the proton, thus probing the proton in nuclear collisions
rather than probing the nucleus in collisions with protons.

\section*{Acknowledgements} We thank Leonid Frankfurt, Vadim Guzey and Daniele Treliani for useful discussions.
The research of MS was partially supported by grant from the US Department of Energy.

\end{document}